\documentclass[aps,preprint]{revtex4}%
\usepackage{amssymb}
\usepackage{amsfonts}
\usepackage{amsmath}
\usepackage{graphicx}%
\providecommand{\U}[1]{\protect\rule{.1in}{.1in}}

\begin{document}
\preprint{ }
\title{BE condensates of weakly interacting bosons in gravity fields}
\author{Yukio Tomozawa}
\affiliation{Michigan Center for Theoretical Physics and Randall Laboratory}
\affiliation{University of Michigan}

\begin{abstract}
The Bose-Einstein (BE) condensates of weakly interacting bosons in a strong
gravity field, such as AGN (Active Galactic Nuclei), BHs (black holes) and
neutron stars, are discussed. Being bound systems in gravity fields, these are
stable reservoirs for the Higgs bosons, and vector bosons of Z$^{0}$ and
W$^{\pm}$ as well as supersymmetric bosons. Upon gravitational disturbances,
such as a gravitational collapse, these objects are relieved from the BE
condensate bound states and decay or interact with each other freely. Using
the repulsive nature of gravity at short distances which was obtained by the
present author as quantum corrections to gravity, the particles produced by
the decays or interactions of the bosons liberated from BE codensates can be
emitted outside the horizon for our observation. It is suggested that the
recently observed gamma ray peak at 129.8 $\pm$ 2.4 GeV from FERMI Large Area
Telescope may be evidence for the existence of the Higgs boson condensates.
The BE condensates of supersymmetric bosons are the most likely sources for
the gamma rays from DMP (dark matter particle) and anti-DMP collisions. It is
shown that the said process from the DMPs spread in the galaxy is too small
for the incident DMP with the intensity of the cosmic ray energy spectrum.

\end{abstract}
\volumeyear{year}

\maketitle

\section{Introduction}

In a recent measurement, Fermi Large Area Telescope has reported a gamma ray
peak at 129.8 $\pm$ 2.4 GeV as a possible evidence for a DMP\cite{fermi}. This
mass value is remarkably close to that of Higgs boson like particles
discovered by LHC experiments\cite{higgs}, with the observed mass value of
126.0 $\pm$ 0.4 GeV or 125.3 $\pm$ 0.4 GeV. If such entirely different
particles have a degenerate mass value, the reason for the degeneracy must be
clarified. If both are identical objects (Higgs boson), then the process for
producing such a gamma ray must be shown. Even if the gamma ray is due to DMP
annihilation, the mechanism for such an encounter of DMP collision is not
clear. Since the density of DMP in our galaxy is not large, even DMP of cosmic
ray intensity cannot produce enough gamma rays due to a small cross section of
weak interactions, as will be explained later. Clearly, a new mechanism or a
new process is required for the explanation. This article provides such a scenario.

\section{BE condensates of Higgs bosons in gravity}

In a strong gravity field of AGN, BHs and neutron stars, BE condensates of
weakly interacting bosons are conceivable. They satisfy Gross-Pitaevskii (PE)
equation\cite{bec},\cite{pit}%
\begin{equation}
\lbrack-\frac{\hslash^{2}}{2m}\nabla^{2}+V(r)+U_{0}\mid\psi(\mathbf{r)\mid
}^{2}]\psi(\mathbf{r)=}\mu\psi(\mathbf{r),} \label{GP}%
\end{equation}
where $m$ and $\mu$ stand for the mass\ and chemical potential of condensate
and%
\begin{equation}
V(r)=-\frac{GmM}{r} \label{gravity}%
\end{equation}
is the gravitational potential of a BH mass, $M$. We consider three
representative cases%
\begin{equation}
M=10^{9}m_{\odot};\text{ }3\ast10^{6}m_{\odot};\text{ \ }3m_{\odot},
\label{mass}%
\end{equation}
representing a typical AGN, the BH at the center of our Milky Way galaxy and a
BH of the smallest mass in our galaxy, respectively. In the case of a neutron
star, the result is close to the third case. In the following, the numerical
estimate for these three cases sre presented, whenever three numbers are listed.

Since a BE condensate is a bound state, all the BE states are stable unless an
extra energy or chemical potential is provided. This is similar to neutrons in
nuclei and neutron stars which\ do not decay. This is a remarkable property of
BE condensates in gravity. Most bosons are unstable in free state, but stable
in a BE condensate state. In a gravitational collapse, they are released to a
free state and decay or interact with each other. This is the essential point
of this article. The emission of the observable events through the horizon of
a BH will be discussed in the next section.

For the boson, we select the Higgs boson with the mass%
\begin{equation}
m_{H}=126GeV
\end{equation}
and its Compton wave length,%
\begin{equation}
\frac{\hslash}{m_{H}c}=1.57\ast10^{-16}cm
\end{equation}
as a representative particle. For the other particles, such as vector bosons,
Z$^{0}$ and W$^{\pm}$, and supersymmetric bosons, one has to change the
numerical values depending on their mass values. \ 

While the horizen scales are%
\begin{equation}
L=\frac{2GM}{c^{2}}=3\ast10^{14}cm;\text{ }9\ast10^{11}cm;\text{ }9\ast
10^{5}cm,
\end{equation}
the magnitudes of the coupling strengths of gravity are%
\begin{equation}
\frac{Gm_{H}M}{\hslash c}=\frac{1}{2}\frac{m_{H}c}{\hslash}\frac{2GM}{c^{2}%
}=0.955\ast10^{30};\text{ }2.87\ast10^{27};\text{ }2.87\ast10^{21},
\end{equation}
and the Hawking temperatures are%
\begin{equation}
T_{H}=\frac{\hslash c^{3}}{8\pi GMk_{B}}=\frac{6.17\ast10^{-8}K}{(M/M_{\odot
})}=6.17\ast10^{-17}K;\text{ }2.06\ast10^{-14}K;\text{ }2.06\ast10^{-8}K.
\end{equation}

At low temperature, all condensates are in the ground states, which reside at
the Bohr radius,%
\begin{equation}
\frac{\hslash}{m_{H}c}\frac{\hslash c}{Gm_{H}M}=1.64\ast10^{-46}cm;\text{
}5.47\ast10^{-44}cm;\text{ }5.47\ast10^{-38}cm.
\end{equation}
This means that the condensates are in a relativistic regime. Also these
distances are inside the Planck scale ($10^{-33}cm$). The physics in that
region is not yet established. Nevertheless, one can estimate the number of BE
condensates, $N$, \ by%
\begin{equation}
\frac{Gm_{H}M}{r}>\frac{\lambda N}{L},
\end{equation}
where $\lambda$ is the coefficient of $\phi^{4}$ term in the Lagrangian,
$\phi$ and $v$ being the Higgs field and its vacuum value, hence%
\begin{equation}
\lambda=(\frac{m_{H}}{v})^{2}=(\frac{126}{246})^{2}=0.26
\end{equation}
Then, choosing r to be the Bohr radius of the BE condensate one gets%
\begin{equation}
N<\frac{L}{r}\frac{Gm_{H}M}{\lambda}=6.72\ast10^{90};\text{ }1.82\ast
10^{83};\text{ }1.82\ast10^{65}.
\end{equation}

For the modification of physics, let us start with the quantum corrections on
gravity, Eq. (\ref{gravity}). The quantum corrections on gravity has been
established recently by the present author\cite{qc}. In this work, the metric
in a spherically symmetric and static metric is given by%
\begin{equation}
g_{00}=g_{rr}^{-1}=1+\frac{r^{2}}{\xi}-\sqrt{\frac{r^{4}}{\xi^{2}}+\frac
{4GMr}{\xi}}, \label{qc}%
\end{equation}
where%
\begin{equation}
\xi=16\pi G\kappa
\end{equation}
and%
\begin{equation}
\kappa=\frac{1}{2880\pi^{2}}(\frac{3}{4}N_{S}+\frac{19}{2}N_{\nu}+\frac
{133}{2}N_{V}),
\end{equation}
$N_{S}$, $N_{\nu}$ and $N_{V}$ being the numbers of scalar fields, four
component neutrino fields and vector fields respectively. For $N_{S}$ = 1,
$N_{\nu}$ = 3 and $N_{V}$ =12, $\kappa$ = 0.0291. Using%
\begin{equation}
\frac{\hslash G}{c^{3}}=2.61\ast10^{-66}cm^{2},
\end{equation}
one gets%
\begin{equation}
\xi=3.82\ast10^{-66}cm^{2}.
\end{equation}
The minimum of Eq. (\ref{qc}) is reached at%
\begin{equation}
r_{0}=(\frac{GM\xi}{2})^{1/3}=6.59\ast10^{-18}cm;\text{ }0.951\ast
10^{-18}cm;\text{ }0.951\ast10^{-20}cm \label{min}%
\end{equation}
with the minimum%
\begin{equation}
(\frac{g_{00}}{2})_{\min}=-(\frac{(GM)^{2}}{4\xi})^{1/3}=-1.14\ast
10^{31};\text{ }-2.37\ast10^{29};\text{ }-2.37\ast10^{25}.
\end{equation}
This value of the minimum, $r_{0}$, indicates that this system is in the range
of relativistic quantum physics. The Lorentz factor at $r_{0\text{ }}$is
estimated as%
\begin{equation}
\gamma=\frac{\hslash}{m_{H}c}\frac{1}{r_{0}}=2.39\ast10^{3};\text{ }%
1.65\ast10^{2};\text{ }1.65\ast10^{4}.
\end{equation}
The lifetime of the Higgs boson in a free state is estimated to be%
\begin{equation}
\tau_{H}=\frac{1}{4MeV}=1.65\ast10^{-22}s
\end{equation}
in the standard model\cite{ilc}. Then the lifetime in BH is given as%
\begin{equation}
\tau_{H}\ast\gamma=3.94\ast10^{-19}s;\text{ }2.72\ast10^{-20}s;\text{
}2.72\ast10^{-18}s, \label{lifetime}%
\end{equation}
if it\ is in a free state.

Replacing the gravitational potential by m*($g_{00}/2$)$_{\min}$, one gets the
upper bound for the Higgs boson condensates%
\begin{equation}
N<\frac{Lm_{H}\mid(g_{00})_{\min}\mid}{2\lambda}=0.838\ast10^{62};\text{
}5.23\ast10^{57};\text{ }5.23\ast10^{47}. \label{max}%
\end{equation}
From the computation of the cross section for a Higgs pair production by gamma
pair\cite{higgspair} in the standard model and from the phase space
consideration, one can estimate the cross section of the reverse reaction%
\begin{equation}
\sigma_{\gamma\gamma}=\sigma(Higgs+Higgs->\gamma+\gamma)=0.004\text{ }%
fb=4\ast10^{-42}cm^{2}, \label{gammacr}%
\end{equation}
for the Higgs bosons at rest. The cross section at higher energy increases
\ linearly with energy E. The gamma rays produced in the volume%
\begin{equation}
V=\frac{4\pi}{3}L^{3}=1.13\ast10^{44}cm^{3};\text{ }3.05\ast10^{36}%
cm^{3};3.05\ast10^{18}cm^{3},
\end{equation}
are%
\begin{equation}
n_{\gamma}=2\ast\frac{1}{2}(\frac{N}{V})^{2}\sigma_{\gamma\gamma}%
cV=7.46\ast10^{48}s^{-1};\text{ }1.07\ast10^{48}s^{-1};\text{ }1.07\ast
10^{46}s^{-1}.
\end{equation}
Using the distance of%
\begin{equation}
R=100Mpc=3.09\ast10^{24}m
\end{equation}
for AGN and%
\begin{equation}
R=10kpc=3.09\ast10^{20}m
\end{equation}
for the galactic BHs (the galactic center and other galactic BH), one gets%
\begin{equation}
\frac{n_{\gamma}}{4\pi R^{2}}=0.0602m^{-2}s^{-1};\text{ }8.92\ast10^{5}%
m^{-2}s^{-1};\text{ }8.92\ast10^{3}m^{-2}s^{-1}.
\end{equation}
Note that the Fermi Large Area Telescope has the area size of 0.7 $m^{2}$. The
time integrated gamma rays that can be observed is then%
\begin{equation}
N_{\gamma}=\frac{n_{\gamma}}{4\pi R^{2}}\frac{L}{c}=6.02\ast10^{2}%
m^{-2};\text{ }2.68\ast10^{7}m^{-2};\text{ }0.268m^{-2}. \label{N}%
\end{equation}
assuming that all gamma rays produced are observed. If the estimate of the
gamma ray cross section, Eq. \ (\ref{gammacr}), is changed, these values are
changed accordingly, but the relative values of the three cases are unchanged.
One should notice that these are the maximum number possible for one
explosion. The accummulation of the BE condensates could be smaller than these
numbers. Also one may not detect all photons due to possible dead time of the
detecting instrument.

The collision time of Eq. (\ref{gammacr}) is estimated as%
\begin{equation}
\tau_{\gamma\gamma}=(\frac{1}{2}(\frac{N}{V})^{2}\sigma_{\gamma\gamma}%
cV)^{-1}=4.90\ast10^{-49}s;\text{ }4.66\ast10^{-49}s;\text{ }4.66\ast
10^{-47}s.
\end{equation}
Now the estimate of Eq. (\ref{N}) should be corrected by the area in which the
Higgs condensates are most likely liberated to decay and interact with each
other. By a gravitational collapse or a collision of binary black holes the
Higgs condensates are sent to the edge of the horizon where the strength of
gravity is weaker. So, most likely the condensates interact at the edge of the
horizen. Let us assume that the region where the condensates are liberated to
interact is between $L_{1}$ and $L$, then the time integrated gamma rays, Eq.
(\ref{N}), must be multiplied by a factor%
\begin{equation}
a=\frac{L-L_{1}}{L}/\frac{V-V_{1}}{V}=\frac{1-L_{1}/L}{1-(L_{1}/L)^{3}}%
=\frac{1}{1+L_{1}/L+(L_{1}/L)^{2}}.
\end{equation}
This factor $a$ approaches 1/3 in the limit of $L_{1}$ approaching to $L$.
Thus the numbers in Eq. (\ref{N}) should be multiplied by a factor of 1/3 for
this correction%
\begin{equation}
aN_{\gamma}=2.01\ast10^{2}m^{-2};\text{ }0.893\ast10^{7}m^{-2};\text{
}0.0893m^{-2}. \label{aN}%
\end{equation}
This shows that the BH in the galactic center is the most efficient emitter of
the gamma rays from a Higgs boson collision, even only a fraction of Eq.
(\ref{N}) is emitted when a neighboring star is gravitationally collapsed to
the central BH. These gamma rays reach at the observers on the Earth in a
short time interval, depending on the size of $L-L_{1}$. The explosion after a
gravitational collapse might repeat itself many times. Each time the BE
condensates are created in the collapsed state. In other words, there could be
an oscillation after a gravitational collapse, and each
oscillation\cite{oscill} can produce the BE condensates and the resulting
gamma rays by the Higgs boson collisions.

Finally, a relativistic correction to the GP equation, Eq. (\ref{GP}), should
be mentioned. One way to accomplish this is to use the Bethe Salpeter
equation\cite{BSE}. In this approach, the interaction terms can be used as
pseudopotential form, so that the argument in this article can be used in that
formalism\cite{BSE2}. For a general relativistic approach, see a
reference.\cite{gr}. An alternative method is to use Dirac equation and\ to
identify the spin average of the solution to the scalar boson field. In this
approach the interaction terms can be carried to a relativistic formulation
and the discussion of this section can be utilized.

\section{Penrose diagram and the ansatz of maximum symmetry}

How can we observe gamma rays produced inside a horizon of a BH? First, the
metric derived by quantum corrections, Eq. (\ref{qc}), behaves%
\begin{equation}
g_{00}=1-2\sqrt{\frac{MG}{\xi}r}. \label{repuls}%
\end{equation}
near the origin. It is a repulsive force at the origin. In such a case, one
has to use a Penrose diagram in order to describe a motion near the horizon,
as in the case of \ the Reissner-Nordstrom metric for a charged BH. See Fig.
34.4 in the reference\cite{grav}. The metric $g_{00}$ vanishes at the outer
horizon,%
\begin{equation}
r_{+}=L,
\end{equation}
and at the inner horizon,%
\begin{equation}
r_{-}=\frac{\xi}{2L}=6.35\ast10^{-81}cm;\text{ }2.12\ast10^{-78}cm;\text{
}2.12\ast10^{-72}cm.
\end{equation}
This value of the inner horizon, $r_{-}$, is much smaller than the Planck
distance ($10^{-33}cm$) and then it may be changed by a theory in the future.
However, one should notice that the repulsive force of the gravity starts at
the minimum of $g_{00}$ at $r_{0}$, Eq. (\ref{min}), which is a quantum
mechanical distance. Besides, Boulware and Deser obtained the same metric, Eq.
(\ref{qc}), independently from the present author for a solution of a string
theory model\cite{string}. In other words, the discussion in this paper might
survive in the realm of a new theory in the future. In fact, the solution of
Eq. (\ref{qc}) is the outcome of the Gauss Bonnet term in general relativity,
a quadratic form of the curvature, and appears in a quantum correction of
field theory as well as in a string theory.

From Fig. 34.4 in ref.\cite{grav}, the Penrose diagram has a multi-universe
structure. Consider a test particle in one of the universes. It falls into the
inside of the horizen. It takes an infinite time in the coordinate time to
cross the horizen, but it requires a finite time in the proper time of the
test particle. After crossing the outer horizon at $r_{+}$, the particle goes
inside the inner horizon, $r_{-}$, and comes out at $r_{-}$ and then crosses
at $r_{+}$ of the next universe. Then, what is the implication of this
phenomenon? One loses the symmetry between the starting universe and the next
universe in this process. If one proposes that a test particle must be
prepared at every universe as copies, then all universes are copies of the
starting universe including the test particle. When a test particle of one
universe comes out at the next universe, all the test particles behave the
same motion and the symmetry of all the universes are restored. This is the
ansatz of maximum symmetry. The advantage of this ansatz is the following:
When a test particle comes out at the next universe, the center of BH in the
next universe is moving according to the two body motion in this ansatz, since
every universe has a test particle partner. When a test particle comes out at
the horizon of the next universe, a copy of the test particle of the previous
universe comes out in the original universe. This is a consistent picture for
all multiple universes. Simply put, this implies that a test particle can
cross the horizon. In the next section, the observed phenomenon of SN87A and
the proposed emission of cosmic rays from AGN can be understood from the
quantum mechanical metric and the ansatz of maximum symmetry for the Penrose diagram.

With this ansatz, one has to change the old concept of BH in which it was
considered that nothing can come out of a BH. With the Penrose diagram for the
quantum mechanical metric, Eq. (\ref{qc}), and the ansatz of maximum symmetry,
all particles liberated from the BE condensates by decays and collisions
should come out from the horizon. See more discussions in ref. \cite{cr4}

\section{The explosion of SN87A and high energy cosmic rays from AGN}

The discussion of the previous section yields an explanation for the
observation of SN87A. A large progenitor, Sanduleak -69$^{\circ}$ 202, in the
Magellanic cloud became a supernova in February, 1987. Namely, the explosion
of SN87A was observed at the location of this progenitor. When the core of the
progenitor collapsed, the collapsed objects did not form a neutron star. Since
a neutron star has not been found at the center of SN87A thus far, the
formation of a neutron star is not the reason for the explosion. The formation
of a neutron star after a supernova explosion is an end result of the whole
process. The core collapse must go inside a smaller region, smaller than the
size of a neutron star which is 20 km or probably inside the horizon which is
3 km for 1 solar mass. It is conceivable that the subsequent explosion could
be caused by the repuslive force of Eq. (\ref{repuls}). High energy particles
produced by the repulsive forces must come out from the inside of the horizon
and cause the explosion of the outside materials as a result of a shock wave.
This explains the nuclear components of cosmic rays from supernova.The fact
that one observed the explosion of SN87A at the same location of the
progenitor could be interpreted as a favorable evidence for the ansatz of
maximum symmetry, which was discussed in the previous section. In summary,
when the core collapsed, the most likely collapsed objects went into the
inside the horizon and by explosion it comes out outside the horizon. The
explosion can be explained by the presence of the repulsive force, Eq.
(\ref{repuls}), and the Penrose diagram for the repulsive component of gravity
and the ansatz of maximum symmetry can explain the observation of an explosion
in the same space of\ the progenitor.

The same idea has been used for the 1985 model of high energy cosmic ray
generation by AGN by the present author.\cite{cr1},\cite{cr2},\cite{cr3}.
Recent data of high energy cosmic rays of the Pierre Auger
Observatory\cite{auger} confirmed this prediction. The ansatz of maximum
symmetry was presented in ref. \cite{cr4}. In this model, a gravitational
collapse in a BH results in an explosion by the repulsive force, Eq.
(\ref{repuls}), and high energy particles such as cosmic rays, gamma rays,
neutrinos and DMPs are emitted from the surface of an expanding heat bath.The
emission of the particles outside the horizen can be provided by the nature of
the Penrose diagram and the ansatz of maximum symmetry, as in the case of
SN87A. The expansion rates in a radiation-dominated and a matter-dominated
expansion provide the energy spectra, $E^{-3}$ and $E^{-2.5}$, above and below
the knee energy at 3 PeV, respectively. The knee energy phenomena at 3 PeV
strongly suggests that the existence of a new mass scale at 3 PeV\cite{cion}.
A further prediction of the model is that all high energy particles produced
by AGN have \ the knee energy at 3 PeV for the energy spectra. This prediction
can be tested by the data of neutrinos, gamma rays and DMPs in the near future.

\section{Gamma rays from Fermi Large Area Telescope}

Now let us come back to the gamma ray observed by Fermi Large Area
Telescope\cite{fermi} at 129.8 $\pm$ 2.4 GeV. The data is taken from the
direction of the galactic center. From the estimate of Eq. (\ref{N}), the data
is consistent with the gamma rays from BE condensates of Higgs bosons from the
BH at the galactic center. It is desirable to get more accurate determination
of the peak mass of the gamma rays and the determination of the sources. Of
course, the data depends on the frequency of gravitational collapses in the BH
at the galactic center. One may assume that at the end of a gravitational
collapse in a BH, the accummulation of BE condensates of the weakly
interacting bosons will be renewed by the condensates which moved inward and
condensates created by the gravitational forces at the center. It is important
to establish that the gamma rays observed are created by the BE condensates of
the Higgs bosons. If confirmed, one can see that BHs are useful reservoirs for
stable Higgs bosons. Nature provides a platform for studying the properties of
the fundamental particles, the Higgs bosons and other weakly interacting
bosons in the future.

\section{Gamma rays from DMP collision}

In a supersymmetric theory, the lowest mass state is a candidate for a DMP. It
may not be a boson. In fact, if a gluino or a gaugino is a DMP, it is a
spinor. Then, all unstable excited particles decay to the DMP eventually.
Bosons in a supersymmetric theory may be unstable, but become BE condensates
in gravity as stable particles. When liberated in a gravitational collapse,
they decay to DMP and the resuting DMPs interact with each other and produce
gamma rays. In order to specify an explicit mass value, one has to fix a
specific theory. I will choose the GLMR-RS theory\cite{glmr} as an example.

I will choose the heavyest mass of the GLMR-RS theory to be the mass scale of
the knee energy\cite{family} of 3 PeV,%
\begin{equation}
m_{3/2}=3PeV.
\end{equation}
Then, the lower end of the mass spectrum which are gauginos becoms%
\begin{equation}
M_{1}=8.9\ast10^{-3}m_{3/2}=26.7TeV
\end{equation}%
\begin{equation}
M_{2}=2.7\ast10^{-3}m_{3/2}=8.1TeV
\end{equation}%
\begin{equation}
M_{3}=-2.6\ast10^{-2}m_{3/2}=-78TeV.
\end{equation}
The masses of the bosons in the theory are expected to be at the range of
$m_{3/2}$, 3 PeV. They are unstable. The lowest mass which is a candidate of
DMP is 8.1 TeV for $M_{2}$. In fact, the analysis of the HESS data\cite{hess}
on the gamma ray spectrum from 8 unidentified sources yields a gamma ray
peak\cite{dmp} of 3 $\sigma$ at 7.6 $\pm$ 0.1 TeV. For the rest of the
section, one will assume that the masses of the DMP and the boson B to be%
\begin{equation}
m_{DMP}=8TeV
\end{equation}
and%
\begin{equation}
m_{B}=3PeV,
\end{equation}
respectively. Further one assumes that the simplest decay mode of B to be%
\begin{equation}
B\rightarrow DMP+\nu+\nu^{C},
\end{equation}%
\begin{equation}
B\rightarrow DMP+\gamma+\gamma
\end{equation}
or%
\begin{equation}
B\rightarrow DMP+\mu+\nu,
\end{equation}
depending on the charge states involved. The decay of the B boson may be
dominated by other processes. I have listed the simplest process of three body
decay here from the point of view of detectability from a distance. The three
dody decays are typically distinguished by a bump in the energy spectrum of
neutrinos and gamma rays at the half of the parent mass, due to a triangular
phase space peaking at the maximum energy. This means that the energy spectrum
of gamma rays and neutrinos emitted from a three body decay of a B boson
should show a bump at 1.5 TeV.

The cross section of gamma ray pair production by a DMP pair is assumed to be%
\begin{equation}
\sigma_{\gamma\gamma}(DMP)=\sigma(DMP+antiDMP\rightarrow\gamma+\gamma
)=10^{-42}cm^{2}, \label{dmpgamma}%
\end{equation}
with a similar value of the Higgs pair cross section, Eq. (\ref{gammacr}).

First, let us estimate the rate of gamma ray production by the DMP distributed
in the galaxy hit by a DMP which is emitted from AGN, similar to the emission
of cosmic rays from AGN. Since it is a gravitational acceleration after
gravitational collapse in a BH, the acceleration of the DMP is
conceivable\cite{cr1},\cite{cr2},\cite{cr3}. One may assume that the energy
spectrum of DMP is identical to that of cosmic rays, except that the spectrum
terminated at its mass, 8 TeV. The flux of cosmic rays at energy of 8 TeV is%
\begin{equation}
Flux=10^{-10}(cm^{2}st\text{ }s\text{ }GeV)^{-1}.
\end{equation}
The number density of the nucleons in the Milky Way galaxy of 10$^{12}%
M_{\odot}$, which has a volume of 10 kpc radius and 1 kpc depth, is%
\begin{equation}
n_{p}=10^{12}\frac{2\ast10^{33}g}{1.67\ast10^{-24}g}/\pi(3.09\ast10^{22}%
)^{2}(3.09\ast10^{21})cm^{3}=129\text{ }cm^{-3}.
\end{equation}
Assuming the same mass density for DMP which is distributed in a spherical
symmetric sphere of a radius of 10 kpc, one gets the DMP number density in the
galaxy%
\begin{equation}
n_{DMP}=n_{p}/8000=1.61\ast10^{-2}\text{ }cm^{-3}.
\end{equation}
For simplicity, one assumes that the observer is at the center of the sphere
of the DMP distribution. Then, integrated the incident flux of DMP over all
direction (4$\pi$ steradian) and 1 GeV energy range, one gets the gamma rays
at the observer%
\begin{align}
N_{\gamma}  &  =2\ast n_{DMP}\text{ }\sigma_{\gamma\gamma}(DMP)\text{ }%
4\pi\text{ }Flux\text{ }GeV\text{ }\int\frac{4\pi r^{2}dr}{4\pi r^{2}}\\
&  =2\ast(1.61\ast10^{-2}cm^{-3})\text{ }(10^{-42}cm^{2})\text{ }4\pi\text{
}(10^{-10}s^{-1}cm^{-2})(3.09\ast10^{22}cm)\\
&  =1.24\ast10^{-30}s^{-1}cm^{-2}%
\end{align}
This is an extremely small number. In other words, the gamma rays from the DMP
pair collision cannot come from the DMP distributed in the Milky Way galaxy,
or in any galaxies. It has to come from the BE condensates in a BH.

The value of the gravitational potential is increased by the ratio of the
masses of the supersymmetric boson and the Higgs boson%
\begin{equation}
\frac{m_{B}}{m_{H}}=\frac{3PeV}{126GeV}=2.38\ast10^{4}, \label{mratio}%
\end{equation}
then the maximum number of the BE condensates of the supersymmetric bosons
become, from Eq. (\ref{max})%
\begin{equation}
N<1.99\ast10^{66};\text{ }1.24\ast10^{62};\text{ }1.24\ast10^{52}.
\end{equation}
Here, one assumed the same value of $\lambda$ as that of the Higgs boson for
simplicity. For the estimate of the observable gamma rays on the Earth, Eq.
(\ref{aN}), one has to remembers that the choice of the combination of DMP and
antiDMP gives a factor of 1/2, that the cross section is assumed be 1/4 of the
Higgs cross section, and that the square of the mass ratio, Eq. (\ref{mratio}%
), must be multilied. In other words, a factor of%
\begin{equation}
\frac{1}{8}(\frac{m_{B}}{m_{H}})^{2}=7.08\ast10^{7}%
\end{equation}
must be multiplied for the value of Eq. (\ref{aN}). Hence one gets%
\begin{equation}
aN_{\gamma}=1.42\ast10^{10}m^{-2};\text{ }6.32\ast10^{14}m^{-2};\text{
}6.32\ast10^{6}m^{-2}%
\end{equation}
for the observable gamma rays on the Earth. These numbers indicate that even
if a small fraction of the gamma rays are emitted by a DMP pair collision, it
is still in the range of observability.

\begin{acknowledgments}
The author would like to thank Peter K. Tomozawa for reading the manuscript.
\end{acknowledgments}

\end{document}